# NIOBIUM TO TITANIUM ELECTRON BEAM WELDING FOR SRF CAVITIES*

M. Parise†, D. Passarelli, J. Bernardini, Fermi National Accelerator Laboratory, 60510 Batavia, IL ,USA

*Abstract*

Titanium and niobium are the main materials used for the fabrication of Superconducting Radio Frequency (SRF) cavities. These two metals are usually joined , using various welding techniques, using a third material in between. This contribution focuses on the development of an innovative electron beam welding technique capable of producing a strong bond between these two different materials. Several samples are produced and tested to assess the mechanical strength at room and cryogenic temperature as well as the composition of the resulting welded joint. Also, the first units of the Single Spoke Resonator type 2 (SSR2) cavities for the Proton Improvement Plan-II (PIP-II [1]) have been fabricated joining directly various grades of titanium to niobium and results gathered through the fabrication will be reported.

## INTRODUCTION

SRF cavities used in particle accelerators are, for the most part, manufactured by forming, machining and Electon Beam (EB) or Tungsten Inert Gas (TIG) welding materials such as Niobium, Titanium and Stainless Steel. Niobium, and in particular high purity niobium, is the metal used for the fabrication of the superconducting resonators because of its high transition temperature, high critical magnetic field and its availability in the forms of sheets and rods/tubes. Given the low operating temperatures of the superconducting structures (around 2 - 4 K) and, in case of the need for proper heat dissipation, liquid helium is required. An external vessel (also called helium jacket), joined to the niobium resonator, is therefore needed. The search for vessel's materials that preserve appropriate strength and ductility at cryogenic temperatures and, among other characteristics, are also paramagnetic, commercially available and machinable/weldable sees the titanium as the one mostly used. One of the challenges of the cavity design is the development of the transition from the niobium resonator to the helium tank. Niobium resonators with titanium helium jackets, are historically joined using a third material that is an alloy of the 2: Nb55Ti. This approach increases the number of welded joints, and thus, the cost of the overall project with respect to joining the 2 materials directly together. While designing the new pre-production SSR2 (ppSSR2) cavity [2], [3], [4] for PIP-II it was decided to develop an EB welded joint between high purity niobium and 2 different types of titanium to prove that such joints can be suitably used for the fabrication. Such joints are mechanically characterized at Room Temperature (RT) and at Liquid Nitrogen (LN) temperature. Energy Dispersive X-ray Spectroscopy (EDS) is also used to check if the titanium can diffuse into the niobium as part of the welding process and following heat treatment.

## MATERIALS AND JOINT TYPES

The materials used for the weld development are:

- High purity niobium (Nb) (Residual Resistivity Ratio >300)

- Titanium Grade 2 (Ti Gr.2) (UNS R50400)

- Titanium Grade 5 (Ti Gr.5) (UNS R56400)

The Ti Gr.2 is a material that can be used for the construction of pressure vessel according to the ASME Boiler and Pressure Vessel Code [5]. Its characteristics are well suited for the fabrication of SRF cavities and it was chosen as helium jacket material for the ppSSR2 cavities. The Ti Gr.5 one of the most used titanium alloys and it has superior mechanical strength however is not a code [5] compliant material. The flanges outside the pressure boundary of the ppSSR2 cavities are fabricated using this titanium alloy so that the sealing surfaces of the Ultra High Vacuum connection have improved resistance to wear and require minimal or no maintenance throughout their lifecycle. Welded samples are prepared joining Nb to Ti Gr.2 and Nb to Ti Gr.5. The joint types for both material combinations are 2: butt welded joint 4.0 mm thick obtained welding in a straight line from one side only and a butt welded joint of a 1" Nb rod to a Ti flange (for both titanium grades). These 2 joint types serve to develop the EB welding parameters on the simplest joint configuration possible (straight line on flat sheets) and to be able to cut specimens for tensile testing and analysis at the Scanning Electron Microscope (SEM). The tensile testing is needed to assess the structural performance of the welded joint. The SEM analysis allows to obtain macrographs of the grain structure, and using the EDS technique we are able establish the chemical composition of the joint and its Heat Affected Zone (HAZ). Also, flanges are used to verify the leak tightness and to reproduce the geometry of the real joint as present on the ppSSR2 cavity flanges. A total of 14 sheets 4" x 2.5" (8 made of Ti Gr.2 and 6 of Ti Gr.5) are EB welded to 14 Nb sheets with exact same dimensions. A total of 4 titanium flanges (2 made of Ti Gr.2 and 2 of Ti Gr.5) are EB welded to 4 Nb rods.

---

## SETUP AND WELDED SAMPLES

All the EB welds are performed using a Sciaky 60kV EB welding machine capable to deliver 30 kW. The vacuum level in the welding chamber at the beginning of each weld is better than $10^{-4}$ Torr. Fig. 1 shows the test setup as described for the 2 joint types. Before welding each niobium part is lightly etched using an acid mixture to remove the oxidized layer and bagged in plastic bags backfilled with nitrogen gas to prevent oxidation. The titanium parts are ultrasonically cleaned only.

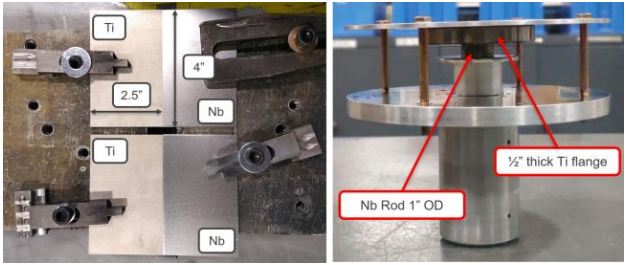

Figure 1: Setup Used for the EB Welds of the 2 Joint Types. On the left: Butt Weld 4.0 mm Thick with Same Base Metal Thickness Welded from 1 Side Only in a Straight Line; on the Right: Butt Weld of the Nb Rod on the Titanium Flange.

A total of 27 welds are performed on the sheets. A total of 4 on the flanges and rods. To begin with, multiple welds are performed on the same welded coupon trying to maximize the number of trials. The parameters used for the joints between Ti Gr.2 and niobium are optimized for a full penetration 4.0 mm thick joint with a smooth underbead and top surface avoiding sputtering on the back side. The parameters used for the joints between Ti Gr.5 and niobium are optimized for maximum penetration (ideally 0.5" deep) as narrow as possible. The two different strategies reflects the joint types on the SSR2 cavity [3], [4]. Fig. 2 shows welds no. 18, 19 and 29 between Nb and Ti Gr.2 and welds 26, 27 and 30 between Nb and Ti Gr.5. Welds 18, 19, 26 and 27 are deemed visually acceptable and satisfying the objectives set for the two different titanium grades. The parameters for the flanges/rods setup are slightly tweaked using the sheets' parameters to achieve full penetration given the different volume of material. Weld 29 is a 4.0 mm deep weld and by examining the back of the flange it is possible to verify that the weld is full penetration. Weld 30 should be 13 mm deep but the weld did not penetrate on the whole circumference.

Table 1 shows some of the parameters used for the welds of Fig. 2. Given the different melting temperature between Nb and titanium, to obtain a symmetric weld, the beam should have an offset toward the metal with the higher melting point (niobium). Oscillation of the beam are used for welds 18, 19 to get a smooth underbead and a Cosmetic Pass (CP) is used to get a smooth top surface.

## TENSILE TESTS

The base materials (Nb, Ti Gr.2 and Ti Gr.5) used to produce the welded coupons are cut according to ASTM standard dimensions [6] into tensile specimens and tested at RT and LN temperature. Tensile specimens are also cut from the welded coupons. Half of all the samples is Heat Treated (HT) in a vacuum furnace at 650 °C for 10 hours while the other half do not go through any heat treatment. Tests are carried out at RT and LN temperature. Average (Av.), maximum and minimum yield and tensile strengths are reported in Table 2 and 3.

Table 1: Weld Parameters Used for the Welded Joints Shown in Fig. 2

| Weld No. | Voltage kV | Current mA | Offset | Oscillation @200Hz |
|---|---|---|---|---|
| 18/19 | 50 | 50 | 0.02" no CP | 0.05"/0.05" 0.16"/0.16" CP |
| 26/27 | 50 | 40 | 0.02" | - |
| 29 | 50 | 165 40 CP | 0.02" | - |
| 30 | 50 | 55 40 CP | 0.02" | 0.05"/0.05" 0.16"/0.16" CP |

Table 2: Tensile Test Results: Yield Strength (MPa)

|  | Av. RT | Max RT | Min RT | Av. LN | Max LN | Min LN |
|---|---|---|---|---|---|---|
| Nb | 76 | 81 | 72 | 578 | 636 | 533 |
| Nb HT | 66 | 70 | 64 | 451 | 552 | 355 |
| Ti Gr.2 | 328 | 333 | 319 | 594 | 613 | 585 |
| Ti Gr.2 HT | 289 | 292 | 286 | 632 | 649 | 615 |
| Ti Gr.5 | 1077 | 1082 | 1072 | 1617 | 1669 | 1565 |
| Ti Gr.5 HT | 1070 | 1077 | 1062 | 1606 | 1696 | 1503 |
| Nb-Ti Gr.2 | 101 | 106 | 93 | 595 | 609 | 579 |
| Nb-Ti Gr.2 HT | 92 | 97 | 87 | 546 | 469 | 586 |
| Nb-Ti Gr.5 | 97 | 98 | 95 | 432 | 440 | 425 |
| Nb-Ti Gr.5 HT | 82 | 86 | 79 | 379 | 398 | 359 |

Table 3: Tensile Test Results: Tensile Strength (MPa)

|  | Av. RT | Max RT | Min RT | Av. LN | Max LN | Min LN |
|---|---|---|---|---|---|---|
| Nb | 191 | 192 | 190 | 674 | 687 | 662 |
| Nb HT | 186 | 189 | 182 | 578 | 593 | 565 |
| Ti Gr.2 | 498 | 505 | 492 | 954 | 958 | 951 |
| Ti Gr.2 HT | 486 | 484 | 488 | 986 | 986 | 986 |
| Ti Gr.5 | 1107 | 1082 | 1072 | 1779 | 1793 | 1765 |
| Ti Gr.5 HT | 1102 | 1109 | 1097 | 1712 | 1737 | 1696 |
| Nb-Ti Gr.2 | 199 | 202 | 197 | 646 | 655 | 637 |
| Nb-Ti Gr.2 HT | 200 | 202 | 195 | 631 | 639 | 625 |
| Nb-Ti Gr.5 | 193 | 194 | 192 | 496 | 508 | 484 |
| Nb-Ti Gr.5 HT | 191 | 192 | 190 | 523 | 558 | 487 |

All welded coupons show yield and tensile strengths (on average) greater than the weaker of the base materials (Nb). The specimens cut from the welded coupons all broke on the

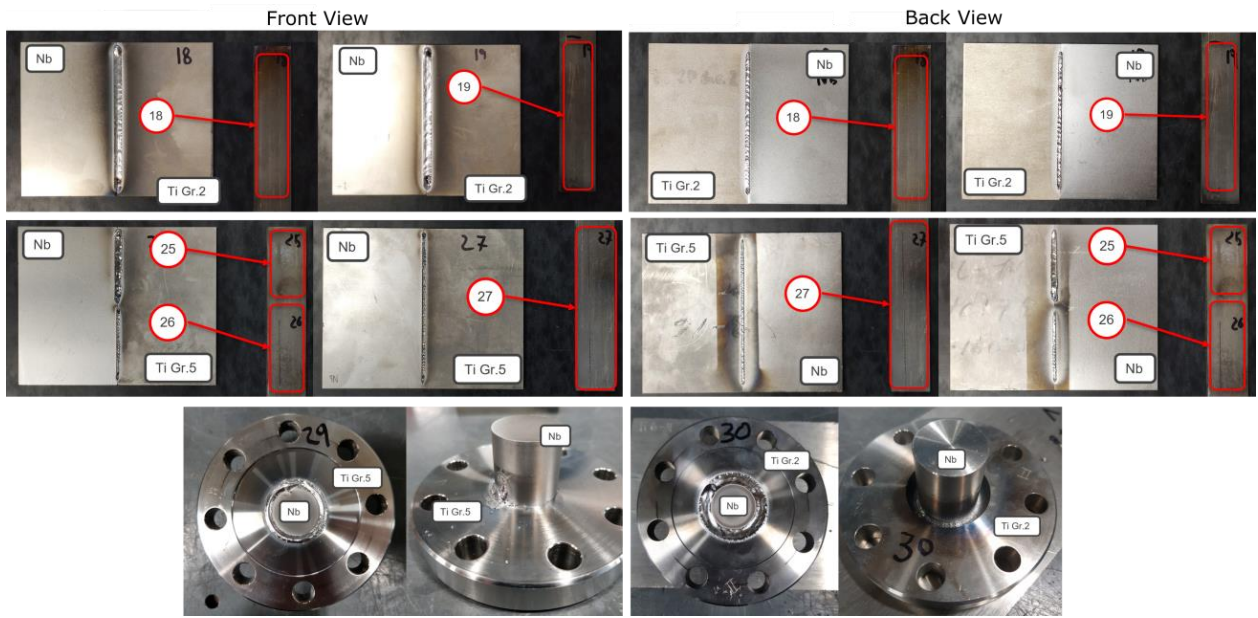

Figure 2: Front and Back of some acceptable welded joints between Nb and Ti Gr.2/Ti Gr.5

Nb side outside the welded area and HAZ both at RT and LN temperature. The Heat Treatment at 650 °C for 10 hours slightly reduce the strength of the base material:Nb [7] and Ti Gr.2. The tensile tests prove that the joints are structurally sound both at RT and LN temperature and HT does not affect the strength of the welded joint more than the base material.

## MACROGRAPHS AND EDS ANALYSIS

Samples are prepared using a diamond saw before mechanically polishing the cross section of the weld. Images are acquired with the usw of a SEM.The EDS elemental characterization is also performed. Fig. 3 shows the cross section of the weld between Nb and Ti Gr.2. On the right the elemental composition can be observed for a selected scan area covering the HAZ on the Nb side. No titanium contamination is measured on the Nb side of the weld for all the samples analyzed. Half of the samples received the heat treatment at 650 °C for 10 hours to verify that the process does not promote titanium diffusion. The heat treatment does not affect the grain size of the HAZ and the morphology of the welded area.

## LEAK CHECK AND COLD SHOCK

A total of 3 out 4 welded flanges go through cleanroom cleaning, which involves ultrasonic cleaning with UPW and detergent, and dry cleaning using ionized nitrogen. The flanges are then sealed with a blanks using the same hardware that will be used on the SSR2 cavities, which includes threaded rods made of 316L Stainless Steel (SS), Silicon Bronze nuts and an aluminum gasket with hexagonal cross section. The blanks used to mate the welded ones are made of 316L SS as the flanges of the bellows that will be assembled on the SSR2 beamline [8]. Sealed flanges are pumped

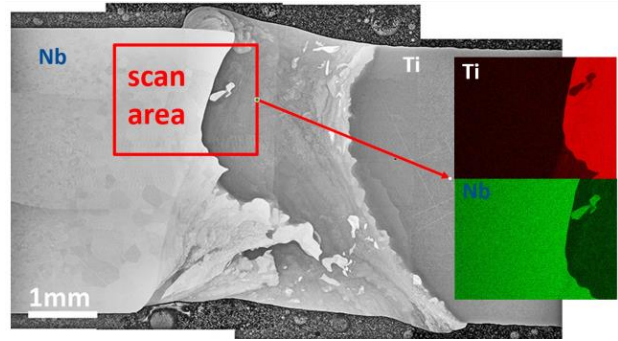

Figure 3: Image acquired with SEM and EDS Analysis of a Nb to Ti Gr.2 Welded Joint. The underbead is on top of the picture and the sample was welded with the underbead looking down as opposed to the orientation of the picture. On the right the EDS elemental characterization ensures no titanium contaminated the niobium side. On the left it is possible to appreciated the HAZ extending for a couple of mm beyond the welded area after that the Nb grain sizes do no show any enlargement compared to the one present in the base material

down and a Helium Mass Spectrometer leak check is used to perform an initial leak check. Subsequently, the sub-assemblies are cold shocked in LN for 10 minutes before another leak check. The whole process is repeated 3 times disassembling the flanges in between and using a new gasket but without changing the SS flanges or the hardware. All flanges passed the leak check before and after any of the cold shocks.

## CONCLUSION

A new type of EB welded joint is developed at Fermilab using Nb and 2 titanium grades. This allows to join directly the SRF cavities' most used materials enabling a step forward in the fabrication and design processes. The joints welded as part of this work showed structural integrity before and after HT and also at LN temperature. Titanium does not appear to contaminate the Nb side of the weld and the HT does not affect its morphology. The welded flanges passed cold shock in LN and leak check. The technology is deemed solid and is implemented for the fabrication of the SSR2 cavities at Fermilab [3], [4].